# Pressure-induced superconductivity in PdTeI with quasi-one-dimensional PdTe chains

Yi Zhao[1#], Jun Hou[2,3#], Yang Fu[4#], Cuiying Pei[1], Jianping Sun,[2,3] Qi Wang[1,5], Lingling Gao[1], Weizheng Cao[1], Changhua Li[1], Shihao Zhu[1], Mingxin Zhang[1], Yulin Chen[1,5,6], Hechang Lei[4*], Jinguang Cheng[2,3*] and Yanpeng Qi[1,5,7*]

1. School of Physical Science and Technology, ShanghaiTech University, Shanghai 201210, China
2. Beijing National Laboratory for Condensed Matter Physics and Institute of Physics, Chinese Academy of Sciences, Beijing, 100190, China
3. School of Physical Sciences, University of Chinese Academy of Sciences, Beijing 100190, China
4. Laboratory for Neutron Scattering, Beijing Key Laboratory of Opto-electronic Functional Materials & Micro-nano Devices, Department of Physics, Renmin University of China, Beijing 100872, China
5. ShanghaiTech Laboratory for Topological Physics, ShanghaiTech University, Shanghai 201210, China
6. Department of Physics, Clarendon Laboratory, University of Oxford, Parks Road, Oxford OX1 3PU, UK
7. Shanghai Key Laboratory of High-resolution Electron Microscopy, ShanghaiTech University, Shanghai 201210, China

# These authors contributed to this work equally.
* Correspondence should be addressed to Y.P.Q. (qiyp@shanghaitech.edu.cn) or J.G.C. (jgcheng@iphy.ac.cn) or H.C.L. (hlei@ruc.edu.cn)

**Abstract**

The quasi-one-dimensional material PdTeI exhibits unusual electronic transport properties at ambient pressure. Here, we systematically investigate both the structural and electronic responses of PdTeI to external pressure, through a combination of electrical transport, synchrotron x-ray diffraction (XRD), and Raman spectroscopy measurements. The charge density wave (CDW) order in PdTeI is fragile and the transition temperature $T_{CDW}$ decreases rapidly with the application of external pressure. The resistivity hump is indiscernible when the pressure is increased to ∼ 1 GPa. Upon further compression, zero resistance is established above ∼ 20 GPa, suggesting the occurrence of superconductivity. Combined XRD and Raman data evidence that the emergence of superconductivity is accompanied by a pressure-induced amorphization of PdTeI.

# 1. Introduction

Both charge-density-wave (CDW) and superconductivity (SC) are two typical collective electronic phenomena, which caused by strong electron-phonon coupling and Fermi surface (FS) instabilities [1-4]. Tuning of the CDW via external parameters like doping [5-7], intercalation [1,8-11] or pressure [12-17] usually lead to discovery of SC. The relationship between CDW and SC has been a subject of extensive investigations over the past decades and complex connections between them have been revealed, including coexistence, cooperative or competition [18-24].

In one-dimensional (1D) system, atomic chains have strong interaction, thus in theory, long-range CDW state hardly exists when the thermal fluctuation is strong at finite temperature. In practice, long-range CDW ordering states exist in some quasi-1D (q1D) systems, e.g. $NbSe_3$ [25,26], $HfTe_3$ [27,28], $K_{0.3}MoO_3$ [29,30], and $TaTe_4$ [31], where 1D chains are embedded in a three-dimensional (3D) structure with weak interchain coupling. Thus, when we modulate the interchain or/and intrachain coupling with high pressure, q1D materials provide a great platform for exploring relationships between different quantum states [32].

Recently, PdTeI with q1D PdTe chains has received much attention since it exhibits unusual electronic transport properties and multiple quantum states. X-ray diffraction (XRD) and neutron powder diffraction studies indicated that there is a dynamic charge separation of Pd ions with local $Pd^{2+}$ and $Pd^{4+}$ pair persisting at high temperature [33]. The long-range CDW transition have been found at $T_{CDW} = 110$ K with CDW vector $q = [0, 0, 0.396(3)]$. Surprisingly, the carrier concentration decreases gradually before $T_{CDW}$, reflecting the existence of strong CDW fluctuation with possible pseudogap state. Moreover, the sliding CDW state appears below $T_2 \sim 6$ K. Thus, PdTeI provides a novel platform for studying the CDW fluctuation and the interplay between CDW and SC states.

High pressure (HP) as a conventional thermodynamic parameter is a pure way with high efficiency in tuning lattice and electronic states, in particular for quantum state. In order to investigate the pressure effect on CDW and explore possible exotic states in PdTeI, we performed experiments of electrical transport, synchrotron XRD as well as

Raman spectroscopy and systematically investigate the electrical transport properties and crystal structure of PdTeI under pressure. We observed that the $T_{CDW}$ is suppressed until 1 GPa and SC emerges up to 3.10 K at 44.5 GPa. In the meanwhile, a metallic amorphous transition has been identified by HP X-ray diffraction and Raman measurements. Our results suggest that the suppression of CDW may be caused by the stability of $Pd^{3+}$ ion and SC could be relevant to the amorphous transition as a result of structural instability arising from the increased coupling of Pd and Te atoms.

## 2. Materials and Methods

High quality single crystals of PdTeI were grown using a hydrothermal method as described elsewhere [34]. Bar-like single crystals with metallic luster used in this work are stable in the air. *In situ* HP resistivity measurements were performed using various apparatus including piston cylinder cell (PCC), palm-type cubic anvil cell (CAC) [35] and diamond anvil cell (DAC). For PCC and CAC, the standard four-probe method was employed with current along the *c* axis. The Daphne 7373 and glycerol were used in PCC and CAC as the pressure transmitting medium. The pressure values in PCC were determined in situ by monitoring the shift of superconducting transition of lead (Pb), while those in CAC were estimated from the low-temperature calibration curve established from separate measurements on the superconducting transition of Pb. It should be noted that the pressure values inside the CAC exhibit slight variations upon cooling, which has been characterized in our previous work [35]. For DAC, four Pt foils were arranged in a van der Pauw four-probe configuration to contact the sample in the chamber for resistivity measurements. A cubic boron nitride and epoxy mixture layer was employed between BeCu gasket and Pt wires as an insulator layer. *In situ* HP XRD measurements were carried out on the beamline BL15U of Shanghai Synchrotron Radiation Facility using x-ray ($\lambda$ = 0.6199 Å). *In situ* HP Raman spectroscopy experiments were performed using a Renishaw Raman spectrometer (laser excitation wavelength $\lambda$ = 532 nm). Pressure was determined by the ruby luminescence method [36].

## 3. Results and discussion

PdTeI crystallizes in a tetragonal structure (Figure 1 (a)) with space group

$P4_2/mmc$ (No. 131). As shown in Figure 1 (b), PdTeI features quasi-1D channels of corner sharing $PdTe_4I_2$ octahedra along the $c$ axis. The channels are connected mutually by the I-I edges of the octahedron of $PdTe_4I_2$ along the $a$ and $c$ axes, there are four Pd-Te chains in each channel. Since the tilt of the octahedron of $PdTe_4I_2$, the Pd-Te chains are not parallel to the $c$ axis straightly. Optical microscope shows a rod-like crystal (inset of Fig. 1(c)). The average compositions were derived from a typical EDX measurement at several points on the crystal, revealing good stoichiometry with the atomic ratio of Pd : Te : I = 31.53% : 33.45% : 35.02% (Figure 1 (c)). At ambient pressure, resistivity measurements on high-quality PdTeI single crystals reveal obvious anomalies $T_{CDW}$ ~ 110K [34], which has been ascribed to the formation of a CDW order as shown in Figure 1 (d).

We carry out a comprehensive HP study on single-crystalline samples in order to investigate the pressure effect on CDW. Figure 2(a) shows the temperature-dependent resistivity $\rho(T)$ of PdTeI single crystals under various pressures up to 7.2 GPa at 0 T. The $T_{CDW}$ can be well defined from the sharp minimum of the $d\rho/dT$ curve as shown in Figure 2(b). But a misalignment may exist in our resistivity measurements with the contributions from both $c$ axis and $b$ axis. In light of the competing nature between CDW and SC, we measured the resistivity $\rho(T)$ of PdTeI under various hydrostatic pressures to further explore whether SC will emerge followed a suppression of CDW by using PCC up to ~ 2.3 GPa and CAC up to 7.2 GPa. With increasing pressure gradually, the hump-like anomaly in $\rho(T)$ and the corresponding minimum in $d\rho/dT$ move to lower temperatures monotonically from ~ 110 K at ambient pressure to ~ 71 K at 0.75 GPa. It should be noted that the $\rho(T)$ curve measured at 0.61 GPa was employed the pressure decreasing process from ~ 2.3 GPa to 0.61 GPa. As shown in Figure 2 (a) and (b), the CDW transition in $\rho(T)$ and $d\rho/dT$ only exhibits very weak feature at 0.75 GPa and cannot be discerned at 1.6 GPa.

As can be seen in the inset of Figure 2 (b), the pressure dependent $T_{CDW}$ determined from the minimum of $d\rho/dT$ shows a complete suppression at ~ 1.3 GPa. However, no SC was observed down to 2 K with pressure further increasing to ~ 2.3 GPa in PCC and 7.2 GPa in CAC. At the pressure of 19.9 GPa, the $\rho(T)$ curve drops to zero at low temperature, suggesting the emergence of SC (Figure 3 (a)). It is clear that the $T_c$ increases monotonously with increasing pressure and up to 3.07 K at 44.5 GPa (Figure 3 (b)). Moreover, the $\rho(T)$ curves as a function of temperature at various fields for 44.5 GPa is shown in Figure 3 (c). When increasing the magnetic field, the resistivity drop is continuously shifted to a lower temperature and no SC is observed at 2.5 T. The temperature dependent upper critical field $\mu_0 H_{c2}(T)$ is shown in Figure 3 (d). Here, the value of $T_c$ is derived from 90% of the normal state resistivity. To determine the upper critical field $\mu_0 H_{c2}(0)$ at 0 K, the Ginzburg–Landau (G-L) formula $\mu_0 H_{c2}(T) = \mu_0 H_{c2}(0)(1 - t^2)/(1 + t^2)$, where t denotes a reduced temperature of $T/T_c$, is used to fit the $\mu_0 H_{c2}(T)$ curves. The obtained $\mu_0 H_{c2}(0)$ is 2.25 T for 44.5 GPa, which is much lower than the Pauli limiting field $H_p(0) = 1.84 T_c = 5.65$ T. It indicates the orbital pair breaking mechanism is dominant in PdTeI.

*In situ* HP synchrotron XRD measurements were carried out on powered single crystals of PdTeI to clarify whether the pressure-induced SC is associated with structural phase transition (Figure 4 (a)). In the low-pressure range, most diffraction peaks of PdTeI could be indexed to the tetragonal $P4_2/mmc$ structure. When increasing the pressure, all peaks slowly shift to higher angles and no structural phase transition is observed up to 13.3 GPa. As shown in Figure 4 (b), both *a*- and *c*-axial lattice parameters decrease with increasing pressure. Interestingly, above 18.22 GPa, apart from the formation of a broad diffusive peak at ≈ 13°, the Bragg peaks disappear from the XRD spectra. It demonstrates that PdTeI may go through an amorphous phase transition persistent up to 75.6 GPa. In addition, upon decompression, the amorphous behavior is maintained. Meanwhile, *in situ* HP Raman spectroscopy experiments were also carried out up to 31 GPa (Figure 4 (c)). At 1.0 GPa, PdTeI displays seven Raman vibrational modes at 34.8, 44.4, 70.4, 81.5, 127.2, 140.8 and 147.4 $cm^{-1}$, respectively. The split of Raman peak around 127 $cm^{-1}$ and the red shift around 141 $cm^{-1}$ (Figure 4 (d)) above about 1.3 GPa may related with the suppressed of CDW. Consistent with the pressure-induced amorphization from the XRD results aforementioned, all the Raman modes disappear above 15.4 GPa corresponding to completion of the structural transition.

On the basis of the above results, we construct a temperature-pressure phase diagram for PdTeI single crystal, as displayed in Figure 5. One can see that the CDW is fragile and $T_{CDW}$ decreases sharply with pressure. By extrapolating this tendency, the CDW transition is estimated to be suppressed completely above around 1 GPa. It should be noted that some typical vibration mode (e.g. 127.2 cm$^{-1}$ and 140.8 cm$^{-1}$ in ambient condition) shows redshift behavior, which may be related to the stability of Pd$^{3+}$ ion [37]. The disappearance of charge separation of Pd ions may be the reason of suppression of CDW under high pressure. Different from many previous report compression can destabilize the CDW and then SC will emerge nearby [13,23,38], here while we do not observe SC around 1 GPa at temperatures down to 1.8 K [39]. With further increasing pressure, SC was observed at around 15 GPa, where a pressure-induced amorphization emerges. The $T_c$ increases with applied pressure and reaches a value of 3.07 K at 44.5 GPa for PdTeI. The transport measurements on different samples for independent runs provide the consistent and reproducible results, confirming this intrinsic SC under pressure (Figure S1). It is very interesting that an amorphous phase of PdTeI could support SC. Similar phenomena were also observed in other materials, such as Pd$_3$P$_2$S$_8$ [40,41], Bi$_4$I$_4$ [42], (NbSe$_4$)$_2$I [43], and (TaSe$_4$)$_2$I [44,45]. This will stimulate further studies from both experimental and theoretical perspectives.

## 4. Conclusion

In summary, we have investigated the electrical transport properties and crystal structures of the q1D CDW material PdTeI using various high apparatus. Experimental results show that the CDW order is suppressed quickly with pressure and disappears above ~ 1 GPa. At higher pressure above 15 GPa, a pressure-induced SC is observed, which is related to a pressure-induced amorphization. Thus, PdTeI provides a novel platform for studying the CDW fluctuation and SC in q1D systems.

**ACKNOWLEDGMENT**

This work was supported by the National Natural Science Foundation of China (Grant No. 12004252, U1932217, 11974246), the National Key R&D Program of China

(Grant No. 2018YFA0704300), Shanghai Science and Technology Plan (Grant No. 21DZ2260400), the authors thank the support from Analytical Instrumentation Center (# SPST-AIC10112914), SPST, ShanghaiTech University. The authors thank the staffs from BL15U1 at Shanghai Synchrotron Radiation Facility for assistance during data collection. The work at IOPCAS was supported by the Beijing Natural Science Foundation (Z190008), the National Natural Science Foundation of China (12025408, 11921004, 11834016, 11904391, 11874400), and the CAS Interdisciplinary Innovation Team. The work at RUC was supported by the Beijing Natural Science Foundation (Grant No. Z200005), the National Key R&D Program of China (Grant No. 2018YFE0202600), the National Natural Science Foundation of China (12274459), and the Beijing National Laboratory for Condensed Matter Physics.

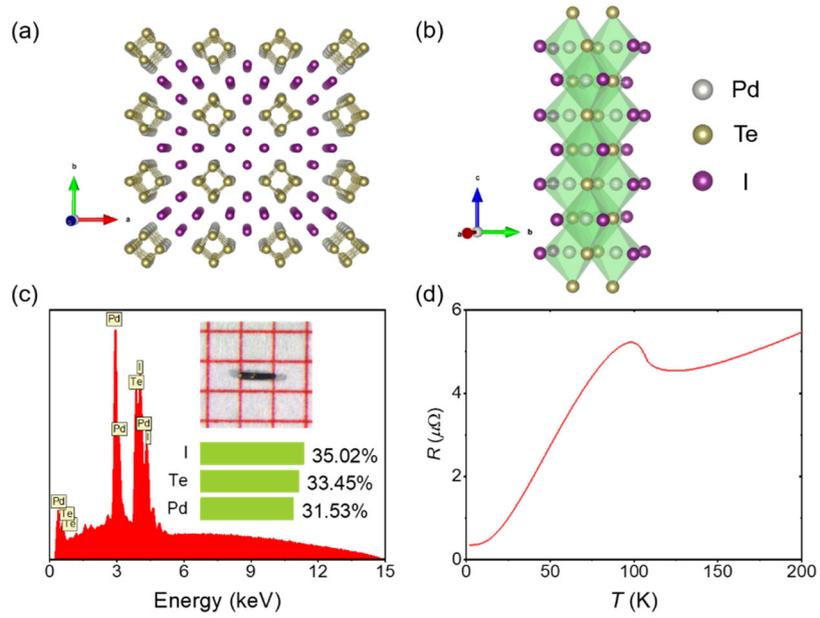

Figure 1. (a) Crystal structure of PdTeI. (b) The PdTe$_4$I$_2$ octahedron highlighted in green, gray, yellow, and violet balls represent Pd, Te, and I atoms, respectively. (c) Energy-dispersive x-ray spectroscopy and optical photograph of PdTeI. (d) Temperature-dependent resistance at ambient pressure for PdTeI single crystal.

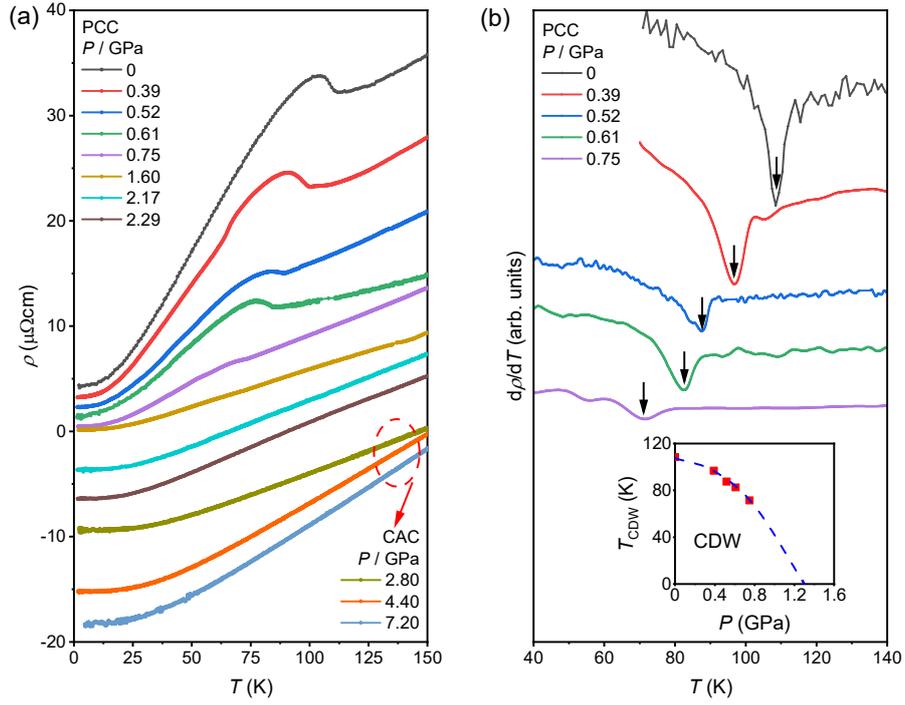

Figure 2. Temperature-dependent (a) resistivity $\rho(T)$ and (b) its derivative $d\rho/dT$ of PdTeI under various hydrostatic pressures up to 7.2 GPa. All the resistivity curves were vertically shifted for clarity. The inset shows the CDW transition temperature $T_{CDW}$ as a function of pressure.

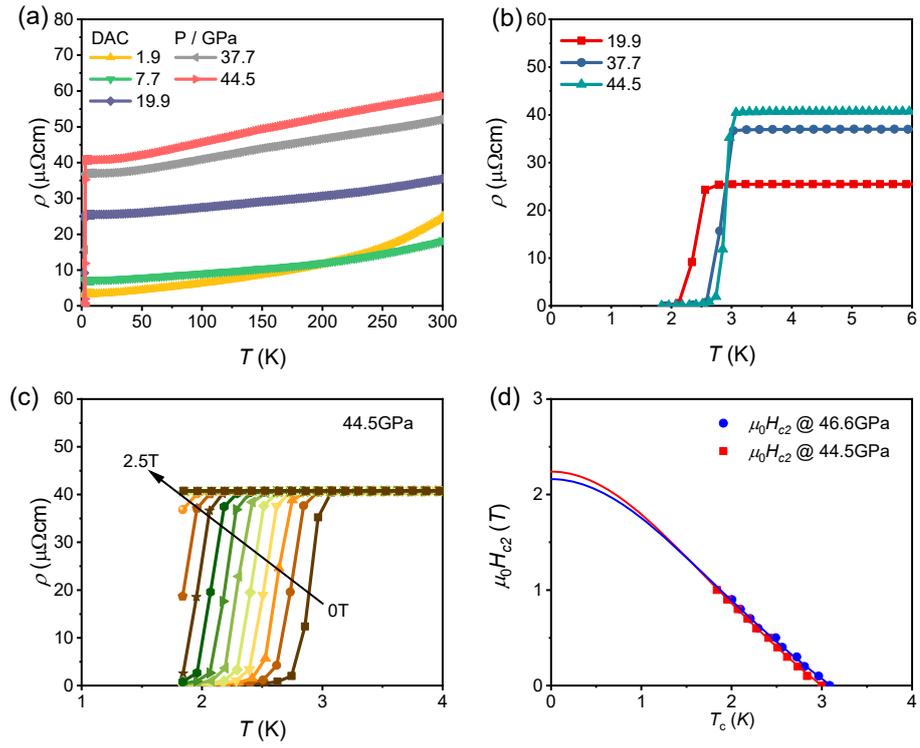

Figure 3. (a) Electrical resistivity of PdTeI as a function of temperature under high pressures up to 44.5 GPa. (b) Temperature-dependent resistivity of PdTeI in the vicinity of the superconducting transition. (c) Temperature dependence of resistivity under different magnetic fields for PdTeI at 44.5 GPa. (d) The upper critical field $\mu_0H_{c2}(T)$ as a function of temperature at representative pressures. The solid lines correspond to the results of fitting by Ginzburg–Landau (G-L) formula.

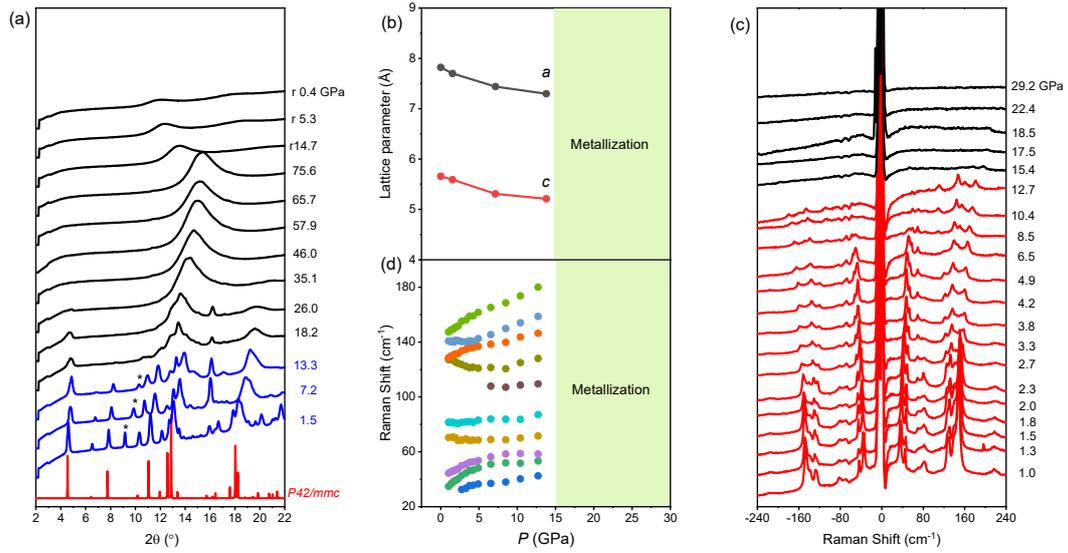

Figure 4. Pressure effect on structure of PdTeI. (a) XRD patterns of PdTeI measured at room temperature with increasing of external pressure up to 75.6 GPa. The x-ray diffraction wave-length $\lambda$ is 0.6199 Å. (b) Pressure dependence of the lattice constants $a$ and $c$ for PdTeI. (c) Raman spectrum of PdTeI at various pressures. Anti-stokes shift and stokes shift of Raman shifts are symmetrical about 0 cm$^{-1}$. (d) Stokes shifts of Raman spectroscopy for PdTeI in compression.

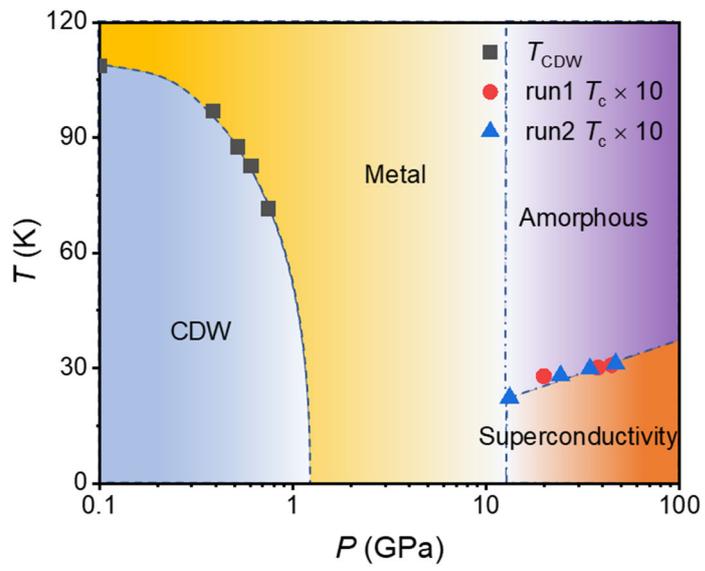

Figure 5. Phase diagram of PdTeI. The $T_{CDW}$ and $T_c$ determined from the resistivity measurements is shown as a function of pressure.
.

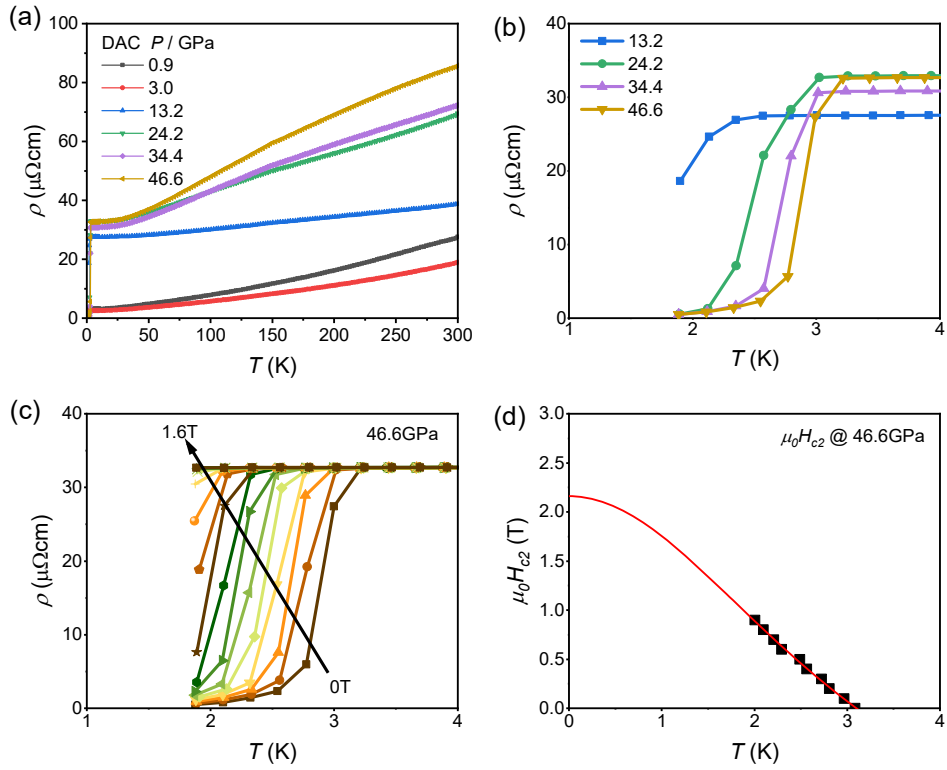

Figure S1. (a) Electrical resistivity $\rho(T)$ of PdTeI as a function of temperature for pressures up to 46.6 GPa in run 2. (b) Enlarged $\rho(T)$ curves in the vicinity of the superconducting transition. Drop of resistivity is obtained for pressure over 13.2 GPa, indicating the emergence of superconductivity. (c) Temperature dependence of resistivity under different magnetic fields for PdTeI at 46.6 GPa in run II. (d) Temperature dependence of upper critical field for PdTeI at 46.6 GPa. Here, $T_c$ is determined as the 90% drop of the normal state resistivity. The solid lines represent the fits based on the Ginzburg–Landau (G-L) formula.